\documentclass[final,onecolumn,prf,aps,amsfonts,amssymb,showpacs,superscriptaddress,floatfix,nofootinbib]{revtex4-1}
\usepackage{amsmath}
\usepackage{graphicx}
\usepackage[dvipsnames]{color}
\usepackage{psfrag}
\usepackage{stmaryrd}
\usepackage{amssymb}
\usepackage{wasysym}
\usepackage{float}
\usepackage{color}

\usepackage[T1]{fontenc}
\usepackage[french,english]{babel}
\usepackage{lmodern}

\usepackage{geometry}
\begin{document}

\title{Space-time resolved measurements of the effect of pinned contact line on the dispersion relation of water waves}

\author{E. Monsalve}
\affiliation{Laboratoire de Physique et M\'{e}canique des Milieux H\'{e}t\'{e}rog\`{e}nes, UMR CNRS 7636, ESPCI-Paris, PSL Research University, Sorbonne Universit\'{e}s, Universit\'{e} Paris Diderot, 10 rue Vauquelin,
75231 Paris CEDEX 5, France}
\affiliation{Laboratoire FAST, UMR CNRS 7608, Universit\'e Paris-Saclay, 91405 Orsay, France}
\author{A. Maurel}
\affiliation{Institut Langevin, UMR CNRS 7587, ESPCI-Paris, 1 rue Jussieu, 75005 Paris, France}
\author{V. Pagneux}
\affiliation{Laboratoire d\'{}Acoustique de l\'{}Universit\'{e} du Maine, UMR CNRS 6613, Avenue Olivier Messiaen,
72085 Le Mans CEDEX 9, France}
\author{P. Petitjeans}
\affiliation{Laboratoire de Physique et M\'{e}canique des Milieux H\'{e}t\'{e}rog\`{e}nes, UMR CNRS 7636, ESPCI-Paris, PSL Research University, Sorbonne Universit\'{e}s, Universit\'{e} Paris Diderot, 10 rue Vauquelin,
75231 Paris CEDEX 5, France}

\date{\today}

\begin{abstract}
We report on an experimental investigation of the propagation of gravity-capillary waves in a narrow channel with a pinned contact line. By using Fourier Transform Profilometry (FTP) we measure the static curved meniscus as well as the surface perturbation. By varying the channel width, between 7 and 15 times the capillary length, we show how edge constraints modify the surface curvature and therefore the dispersion relation. From the space-time resolved field, we obtain a decomposition of the linear mode onto transverse modes satisfying the condition of pinned contact line. This approach, in which we complement the theoretical model with experimental analysis, allows computations of wavenumbers and natural frequencies with a robust statistics. We verify experimentally the convergence of the model and the pertinence of the linear approximation. In addition, we analyze the relative contribution of the experimentally measured static meniscus. An excellent agreement between the computed natural frequencies and the forcing frequency confirms the contribution of the actual space-time resolved measured surface. These experimental results are an accurate estimation of the influence of the additional restoring force exerted by the pinned contact line on the deformed surface which increases the wave celerity. The local character of this effect is evidenced by the decrease of the shift of the dispersion relation as a function of the channel width.
\end{abstract}

\pacs{}
\maketitle

\section{Introduction}

In the dynamics of surface waves, capillary effects become important when the geometry of the container is in the same order of magnitude as the capillary length or in low gravity conditions, where the main restoring force is the surface tension \cite{kopachevskii1972hydrodynamics,zhang2013capillary,berhanu2020capillary}. Several recent applications and experimental works make that the problem of calculating the damping and eigenfrequencies of gravity-capillary waves is still an active subject \cite{shao2021role,viola2018theoretical,horstmann2020linear,ibrahim2015recent}. When edge constraints are added (physical restrictions to the movement of the contact line between the free surface and the container), the change in the dynamic of propagating waves could become non-negligible. In fact, when the wave amplitude is in the order of few millimeters and when the geometry of the container is in the order of few centimeters, the type of boundary condition is crucial \cite{hocking1987damping}. Depending on the wetting conditions and the filling height of the container, the contact line can be pinned in a brimful container \citep{benjamin1979gravity,kidambi2009meniscus}, pinned with a low wettability (tendency of the liquid to be in contact with the solid surface and inversely proportional to the contact angle) \cite{shankar2007frequencies}, slipping with low wettability and surface displacement greater than the slipping threshold \cite{brochard1992dynamics,hocking1987damping,kidambi2009capillary} (see details of slipping threshold in \cite{cocciaro1993experimental}) or slipping with high wettability (hydrophilic boundary) \cite{de2004capillarity,de1985wetting}, in which the effects of edge constraints become negligible, i.e., the motion of the contact line can be considered as free and the dispersion relation of water waves is independent of the size of the container \cite{henderson1990single,wu1984observation,kim2020capillary}. In each one of these boundary conditions the dynamic of the contact line modifies the propagation of surface waves in damping and dispersion. Since wetting is a multi-scale problem, the	 characteristics length and velocity of the phenomenon determine together the mechanism that dominate the phenomenon. Indeed, the surface tension $\sigma$, the density $\rho$ and the gravity acceleration $g$ give the characteristic capillary length $\lambda_c=\sqrt{\sigma/\rho g}$ below which the surface tension is important with respect to the gravity. Besides, the ratio between $\sigma$ and the dynamic water viscosity $\mu$ gives the characteristic velocity $V_c=\sigma/\mu$ below which, as in this work, hydrodynamic losses dominate with respect to molecular features in the wetting process \cite{brochard1992dynamics,de2004capillarity}. 

Several theoretical models have already been developed in the last decades, where some linear approximations \cite{benjamin1979gravity,hocking1987damping,miles1991capillary,henderson1994surface,shankar2007frequencies}, or non-linear approximations \cite{cocciaro1991capillarity, cocciaro1993experimental}, have tried to model the damping and eigenmodes of surface waves with different wetting-conditions. In particular, the linear models proposed by \cite{hocking1987damping,miles1991capillary,mccraney2021resonant} relates, at the lateral boundaries, the vertical velocity and the surface gradient normal to the wall. This linear boundary condition spans from a complete pinned contact line to a moving contact line. Some of the models have simplified the problem by considering a $90^{\circ}$ contact angle (brimful condition) for the static profile \cite{benjamin1979gravity,graham1983new,benjamin1985long, henderson1994surface}. Other models have considered a meniscus with contact angle different than $90^{\circ}$, in which, the inviscid limit has beeen explored in the studies by \cite{nicolas2005effects,shankar2007frequencies,kidambi2009capillary}. Besides, viscous conditions have been considered by \cite{nicolas2002viscous} for the case of $90^{\circ}$ contact angle or by \cite{kidambi2009meniscus} for the case of a circular geometry and concave meniscus with small contact angle. Moreover, an extensive study about the frequency and damping rates of the surface waves modes with pinned contact line in a vertically vibrating container can be found in \cite{howell2000measurements} and recent experimental works have focused on the measurement of wave damping due to the meniscus in non-wetting conditions \cite{michel2016acoustic}. In particular, the model developed by \cite{shankar2007frequencies} focuses on the propagation of progressive waves in a narrow channel where the static meniscus is curved with a small contact angle. This is the case of the present work where we have considered gravity-capillary waves in a rectangular channel propagating in one direction. The laterals walls, parallels to the direction of wave propagation, have a concave meniscus which is a very common case widely used in experimental investigations \cite{monsalve2019perfect,berraquero2013experimental,
bobinski2018backscattering,alarcon2020faraday,huang2020streaming}. 

In this work, we present a direct measurement of the surface displacement and curvature in the whole field, especially close to the lateral walls where capillary effects become important. Our objective is to take advantage of the latest techniques developed in measuring the water surface \cite{cobelli2009global,maurel2009experimental} to verify experimentally the influence of the surface curvature and contact angle in the propagation of small amplitudes waves, in conditions where the Bond number $\mathrm{Bo}=\rho g W^2/\sigma$ and Reynolds number $\mathrm{Re}=\rho \textbf{u}W/\mu$ are small, with $W$ the channel width and $\textbf{u}$ the fluid velocity. The article starts by presenting the theoretical model that allows computation of eigenfrequencies. Then the experimental method and measurements are detailed before comparing, eventually, the experimental dispersion relation to the theoretical model.

\section{Theoretical model}

In this article we revisit the model developed by \cite{shankar2007frequencies} (and references therein) but inserting the experimentally measured values of the static meniscus as well as the transverse profile of the surface displacement from which we obtain a transverse modal decomposition.

\begin{figure}[h]
\centering 
	\includegraphics[width=7.7cm]{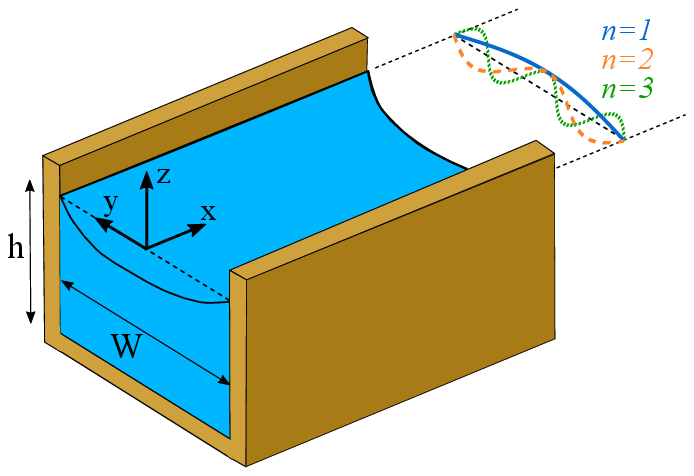}%
 \caption{Coordinate system of the rectangular channel for the propagation of surface waves with pinned contact line}
	\label{fig1}
\end{figure}

Let us consider a water waves rectangular channel of depth $h$ and width $W$. As shown in Fig. \ref{fig1}, we define the $x$-direction or longitudinal as the one of the wave propagation, and the $y$-direction as the transverse one. The coordinate system is located at the center of the channel in the transverse direction ($y$) and at the
 contact line in the vertical direction ($z$). The governing equations at the free surface $z=\hat{\eta}$ are
\begin{align}
\frac{\partial \hat{\eta}}{\partial t} + \left( \textbf{u} \cdot \nabla \right) \hat{\eta} =& u_z, \label{eq:eq1}\\
\frac{\partial \hat{\phi}}{\partial t} + \frac{1}{2}\left( \nabla \hat{\phi} \right)^2 + g\hat{\eta} =& \frac{\sigma}{\rho} \hat{c},
\label{eq:eq2}
\end{align}
where $\textbf{u}=\nabla \hat{\phi}$ is the water velocity, $\hat{c}$ the surface curvature and $\hat{\eta}(x,y,t)$ the instantaneous free surface position. The system satisfies the boundary conditions of impermeable lateral walls $\partial_y \hat{\phi} \left( y=\pm W/2 \right) = 0$ and bottom $ \partial_z \hat{\phi} \left( z=-h \right)=0 $. Besides, the surface is pinned at the contact line, that is, $\hat{\eta}\left(x,y=\pm W/2,t\right)=0$.

The surface displacement  around the still position is produced by small amplitude waves. In this case the still level corresponds to the static meniscus, which we denote $\eta_s$, and is equal to zero at the contact line for simplicity in the calculation $\left[\eta_s\left(x,y=\pm W/2\right)=0\right]$. The linearization of the system is done via formal expansions of the surface displacement, velocity potential and surface curvature,
\begin{align}
\hat{\eta}(x,y,t)=&\eta_s(x,y)+\epsilon \tilde{\eta}(x,y,t) + ..., \label{eq:eta expansion}\\
\hat{\phi}(x,y,z,t)=& \epsilon \tilde{\phi}(x,y,z,t) + ..., \label{eq:phi expansion}\\
\hat{c}(x,y,t)=&c_s(x,y)+\epsilon \tilde{c}(x,y,t)+...,
\label{eq:kappa expansion}
\end{align}
where $\epsilon$ is a small ordering parameter. We insert the above expansions in eqs. \eqref{eq:eq1} and \eqref{eq:eq2} to obtain, at the order $\epsilon^0$, the trivial static solutions $\partial_t \eta_s=0$ and $g\eta_s=\sigma/\rho c_s$, and at the order $\epsilon$, the linearized equations
\begin{align}
\frac{\partial \tilde{\eta}}{\partial t} + \left( \textbf{u} \cdot \nabla \right) \eta_s =& u_z, \label{eq:eq6}\\
\frac{\partial \tilde{\phi}}{\partial t} + g\tilde{\eta}  =& \frac{\sigma}{\rho} \tilde{c},
\label{eq:eq7}
\end{align}
where $\tilde{c}(x,y,t)$ is the three dimensional curvature of the water surface, whose expression is well known and detailed in the appendix \ref{appendixA}. It is worthwhile to mention that the linearized eqs. \eqref{eq:eq6} and \eqref{eq:eq7}, in the case without edge constraints (infinite domain in the plane $(x,y)$) lead to the dispersion relation of gravity-capillary waves
\begin{equation}
\omega^2=\left(g k + \frac{\sigma}{\rho} k^3 \right) \tanh kh,
\label{eq:DR}
\end{equation}
where $k$ is the wavenumber. Considering the harmonic regime we define
\begin{align}
\tilde{\eta}(x,y,t) =& \Re\left[ \eta(x,y) e^{-i\omega t} \right],\label{eq:harmonic_eta} \\
\tilde{\phi}(x,y,z,t) =& \Re\left[ i \phi(x,y,z) e^{-i\omega t} \right],\label{eq:harmonic} \\
\tilde{c}(x,y,z,t) =& \Re \left[ c(x,y) e^{-i\omega t} \right], \label{eq:harmonic_kappa}
\end{align}
where $\eta(x,y)$ and $\phi(x,y,z)$ are functions that satisfy the boundary conditions of impermeable walls and bottom, as well as pinned contact line $\left[\eta(x,y=\pm W/2)=0\right]$. Here and in what follows we have omitted the frequency dependence due to the harmonic regime, thus the theory is developed for a fixed frequency. At the lateral boundaries, the impermeable condition imposes a zero normal velocity ($u_y=0$).  However, a nonzero tangential velocity $u_z$ may exist in inviscid models like this, as written in the right hand of eq. \eqref{eq:eq6}. Thus, considering that we use an inviscid approximation in the dynamics close to a solid surface, i.e., viscous effects in the boundary layer are neglected, the condition of a pinned contact line forces the model to look for \textit{weak} solutions. A theoretical demonstration of the validity of this kind of solution can be found in \citep{benjamin1979gravity,prosperetti2012linear} and references herein.

We expand $\eta(x,y)$ and $\phi(x,y,z)$ as follows
\begin{align}
\eta(x,y)=&\sum_{n=1}^{\infty} A_n(x) \cos\left(\nu_n y\right), \label{eq:eta_projection}\\
\phi(x,y,z)=&\sum_{n=1}^{\infty}  B_n(x) \frac{\cosh \left[\lambda_n (z+h)\right]}{\sinh \left(\lambda_n h\right)} \cos\left(k_n y\right),
\label{eq:phi_projection}
\end{align}
with
\begin{align}
k_n =& 2(n-1) \frac{\pi}{W}, \hspace{20pt} n=1,2,3,...,  \label{eq:k_n} \\
\nu_n =& (2n-1)\frac{\pi}{W}, \hspace{20pt} n=1,2,3,..., \label{eq:nu_n} \\
\lambda_n^2 =& k_n^2+k_x^2, \hspace{20pt} n=1,2,3,...,  \label{eq:lambda_n}
\end{align}
where $k_x$ is the wavenumber in the direction of the wave propagation and is an independent variable (input of the model). For the case of the surface displacement, we present in Fig.  \ref{fig1} the profile of the first three transverse modes of the basis $\cos(\nu_n y)$. The functions $A_n(x)$ and $B_n(x)$ are obtained from the transverse decomposition along the axis of wave propagation. In particular $A_n(x)$ can be written as a superposition of a right- and left-going waves
\begin{equation}
A_n(x)=a_n\left(e^{ik_{x,n}x}+r_n e^{-ik_{x,n}x} \right),
\label{eq:tilde_a_n}
\end{equation}
which we use to fit the coefficients $a_n$, $r_n$ and the wave numbers $k_{x,n}$ (complex valued). Considering that we look for modes that are periodic in the $x$-direction (and in time at the frequency $\omega$), we have that $k_x=k_{x,1}=k_{x,2,}=...=k_{x,n}$. Thus, in the following we shall use only $k_x=k_{x,1}$ as the experimental measured value replaced in eq. \eqref{eq:lambda_n}.

We replace the expression in eqs. \eqref{eq:eta_projection} and \eqref{eq:phi_projection}  in eqs. \eqref{eq:eq6} and \eqref{eq:eq7}, and project the equations onto the basis $\cos \left(\nu_n y\right)$ to get the system of equations
\begin{align}
\frac{\omega}{4}a_i =& \sum_{n=1}^{\infty} \gamma_{i,n} b_n, \hspace*{20pt} i=1,2,3,...,  \label{eq:17a}\\
\omega \sum_{n=1}^{\infty} \delta_{i,n} b_n =& \sum_{n=1}^{\infty} \xi_{i,n} a_n, \hspace*{20pt} i=1,2,3,..., \label{eq:17b}
\end{align}
where the matrices $\gamma_{i,n}$, $\delta_{i,n}$ and $\xi_{i,n}$ represent the integrals detailed in appendix \ref{appendixB}. The system of $2n$ equations \eqref{eq:17a}-\eqref{eq:17b} admits non trivial zero solutions if
\begin{equation}
 \sum_{n=1}^{\infty}\xi_{i,n} a_n = \frac{\omega_i^2}{4} \sum_{n=1}^{\infty} \left[ \sum_{l=1}^{\infty} \delta_{i,l} \gamma_{l,n}^{-1} \right] a_n, \hspace*{20pt} i=1,2,3,...,
\label{eq:20}
\end{equation}
This equation is a matrix eigenvelue problem from which we can obtain the eigenfrequencies $\omega_i$. In practice, we consider only the first eigenfrequency $\omega=\omega_1$ which corresponds to the wave forcing frequency. In this work, considering that the surface displacement $\eta(x,y)$ is measured experimentally, $a_n$ can be obtained directly from the projection of the whole field $\eta(x,y)$ onto the transverse modes basis of eq. \eqref{eq:eta_projection} and from the fit in eq. \eqref{eq:tilde_a_n}.

\section{Experimental set-up}

Waves propagate through a narrow waveguide which, in order to change the channel width, consists of movable walls. A piston type wavemaker, driven by a linear motor (LinMot P10-70) covering the whole water depth and adapted to each specific width, generates waves in the frequency range $\omega=2\pi f \in [2\pi,10\pi]$ $\mathrm{s}^{-1}$ with a precise frequency step of 0.1 Hz between experiments. In each experiment, carried out in a harmonic regime (fixed $\omega$), the motion of the wave maker is sinusoidal in the form $X(t)=A_{\mathrm{wm}}\sin (\omega t)$, with $A_{\mathrm{wm}} \in [0.3, 5]$ mm. 
\begin{figure}[h]
\centering 
	\includegraphics[width=12cm]{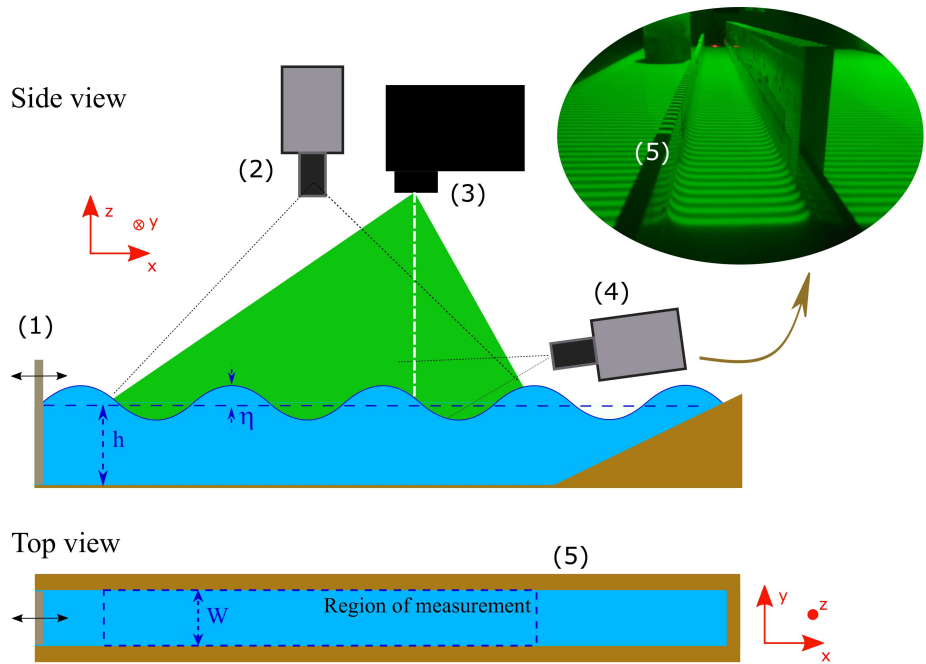}%
 \caption{Experimental set-up: A piston-type wavemaker (1) generates waves that propagate in a waveguide. A fast camera (2) and a video projector (3) are used to measure waves by using FTP technique. An additional camera (4) is located in front of the wavemaker to record the transverse profile of the surface. In the inset, a view of the front camera shows the movable waveguides (5) and the fringes projected onto the water surface.}
	\label{fig1photo}
\end{figure}
As shown in Fig. \ref{fig1photo}, the water channel is $h=50$ mm deep, $L=1$ m long and the width varied between $W= [22,32,42]$ mm. The movement of the wavemaker was previously calibrated in $A_{\mathrm{wm}}$ vs $\omega$ in order to keep the wave amplitude at $a=0.4$ mm in all the frequency range. This wave amplitude is well below the capillary length $\lambda_c=\sqrt{\frac{\sigma}{\rho g}}=2.7$ mm. At the end of the channel an absorbing beach of 10\% slope avoids spurious wave reflections. The deformation of the water surface was measured using the FTP technique \cite{cobelli2009global,maurel2009experimental}, which measures the displacement of a pattern projected onto a diffusively  reflective surface. The pattern is composed of fringes  with a sinusoidal variation in light intensity. The fringes are projected perpendicular to the waveguides, being the phase gradient of the sinusoidal variation parallel to the direction of the wave propagation ($x$-direction). The phase shift between a reference surface and a deformed surface gives, via an optical relation \cite{takeda1983fourier}, the surface height in each pixel of the image. The spatial resolution was $dx=dy=0.7$ mm and the recording frame rate was $f_s=50$ fps. As we observe in Fig. \ref{fig1photo}, the sinusoidal pattern is projected in the $y$-direction. Thus, the classical filtering of the carrier mode of the projected pattern was computed in the $x$-direction \cite{takeda1983fourier}. Instead, no filtering was applied in the $y$-direction. For each experiment, FTP acquisition covers 12 s starting 1 min after the wavemaker to avoid the transient part and reach a stationary regime. In order to get a pinned contact line, experimental trials showed that the surface of a plastic (PVC) wall, when is cleaned with ethanol, gives uniform hydrophobic conditions which helps the contact line to stay pinned subject to the wave perturbations. Thus, before the experiments, the waveguides were previously treated with ethanol and then submerged slowly in still water in order to avoid wetting the zone above the contact line. For comparison, we have also set hydrophilic conditions by covering the lateral walls with a nylon wire-mesh with an opening of $0.1$ mm. The wire-mesh was previously wet in order to have a hemiwicking state \cite{kim2016dynamics}.

\section{Experimental results}
\subsection{Qualitative observations}

\begin{figure}[h]
\centering  
	\includegraphics[width=10cm]{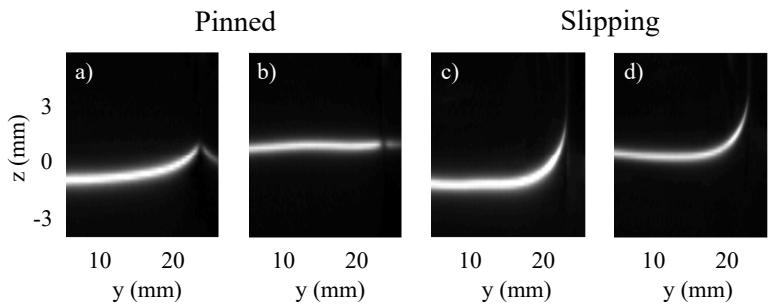}%
 \caption{Transverse profile in the meniscus zone for $\omega=4\pi$ $\mathrm{s}^{-1}$ and $W=42$ mm. $a)$ Wave trough with pinned contact line; $b)$ Wave crest with pinned contact line; $c)$ Wave trough with slipping contact line; $d)$ Wave crest with slipping contact line. In this case $z=0$ corresponds to the still water level without waveguides.}
	\label{fig2menisc}
\end{figure}

To begin, in order to get some qualitative observations, we carried out visualizations of the transverse profile in the plane $(y,z)$ by projecting a white line over an opaque surface (the water has been previously colored with titanium dioxide which does not modify the surface tension nor the wave damping as was shown in the comprehensive study by \cite{przadka2012fourier}). A fast camera located at the end of the channel (see Fig. \ref{fig1photo}) recorded the transverse profile when waves pass through the projected line. In Fig. \ref{fig2menisc} we show snapshots of the meniscus profile.  For comparison, different boundary conditions were tested. In Figs. \ref{fig2menisc}$a$ and \ref{fig2menisc}$b$ the waveguide has a bare face (hydrophobic) which pins the contact line during the wave induced motion. In contrast, in Figs. \ref{fig2menisc}$c$ and \ref{fig2menisc}$d$ a slipping contact line is imposed by using a wire mesh that keeps the wall wet. On both cases, left panels (Figs. \ref{fig2menisc}$a$ and \ref{fig2menisc}$c$) correspond to a wave trough and  right panels (Figs. \ref{fig2menisc}$b$ and \ref{fig2menisc}$d$) correspond to a wave crest. 
	
\subsection{Quantitative results}

\begin{figure}[h]
\centering 
	\includegraphics[width=9cm]{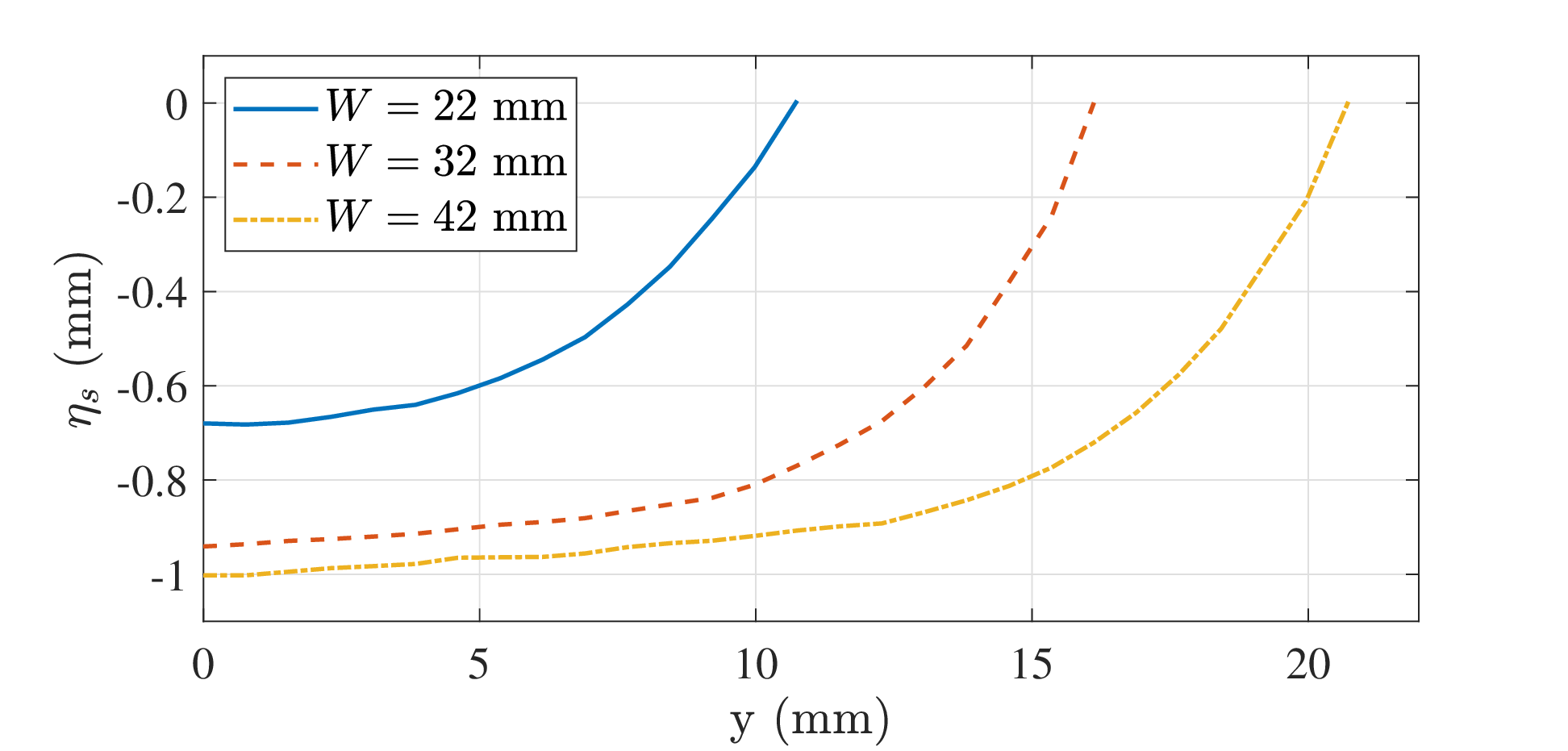}
 \caption{Static meniscus measured experimentally for three different channel widths. The vertical axis is shifted to set the contact line at $z=0$.}
 	\label{fig4etaStatic}
\end{figure}

The static profile $\eta_s$ was measured using FTP by taking as a reference image the water surface without waveguides, it is, completely flat. Thus, the still water surface, deformed by the static meniscus was measured in the whole field, being as expected, invariant along the longitudinal direction $x$. An example is shown in Fig. \ref{fig4etaStatic}, where the static profile for the three channel widths are compared. The vertical axis in Fig. \ref{fig4etaStatic} was shifted in order to set the contact line at $z=0$. We observe how the curvature of the static meniscus changes the water level in the center of the channel ($y=0$), especially for the narrowest case. 

\begin{figure}[h]
\centering 
	\includegraphics[width=14cm]{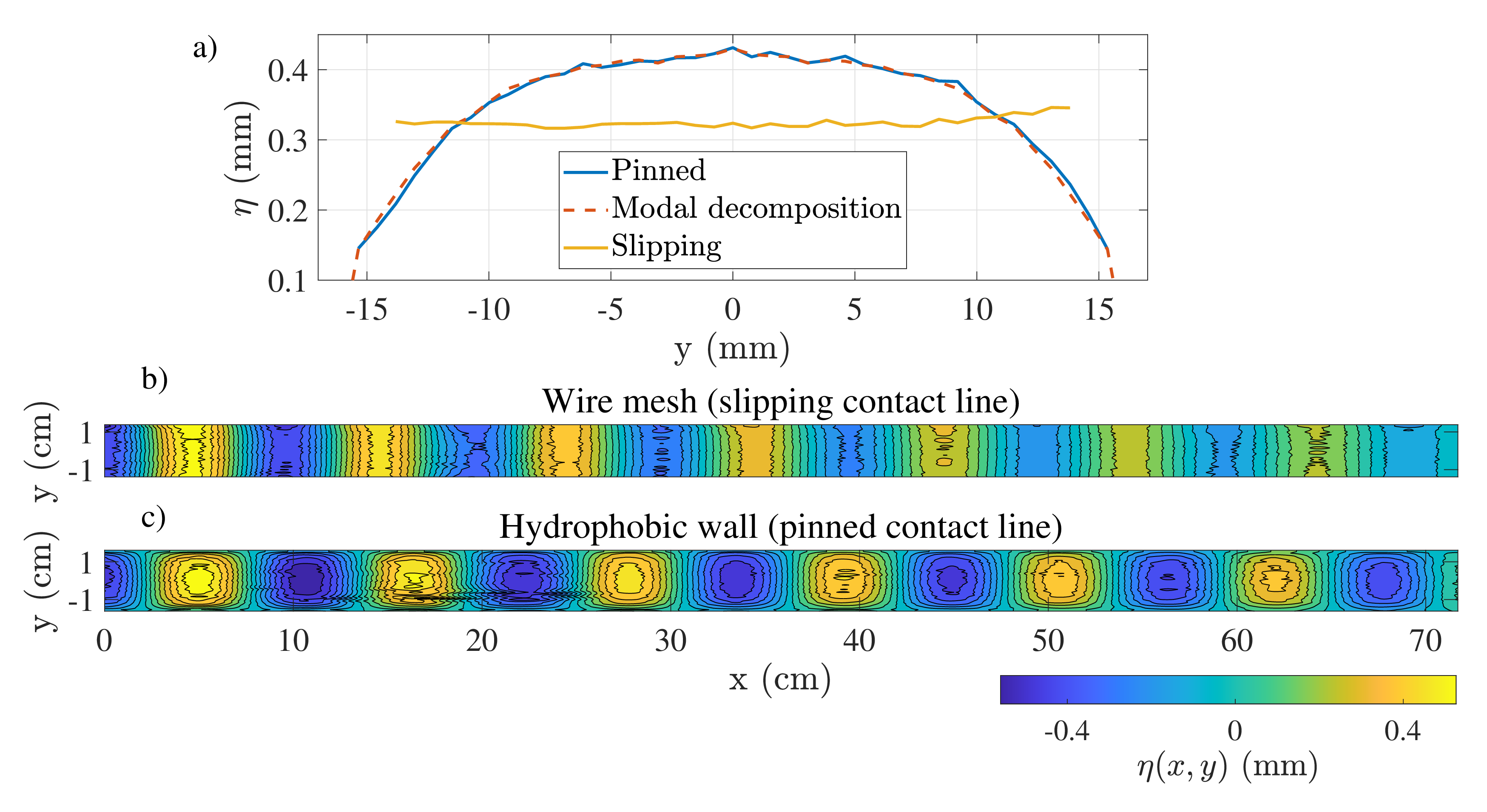}%
 \caption{FTP measurements of the propagation of waves with slipping and pinned contact line conditions. $a)$ Transverse profile of the surface perturbation ($\omega=8\pi$ $\mathrm{s}^{-1}$; $W=32$ mm; $x=0.6$ m). $b)$ Wave field with slipping contact line ($\omega=8\pi$ $\mathrm{s}^{-1}$; $W=32$ mm; Wavemaker phase $\omega t=3\pi/2$); $c)$ Wave field with pinned contact line ($\omega=8\pi$ $\mathrm{s}^{-1}$; $W=32$ mm; Wavemaker phase $\omega t=3\pi/2$).}
	\label{fig5pcolor}
\end{figure}

Quantitative experiments were performed via the optical FTP technique. We measured the space-time resolved surface perturbation $\eta(x,y,t)$ from which we extract the linear mode from a temporal Fourier decomposition
\begin{equation}
\eta\left( x,y,\omega \right)=\frac{1}{T} \int_0^T \eta (x,y,t) e^{i\omega t} dt, 
\label{eq:Fourier_decomp}
\end{equation}
where $T=2N\pi/\omega$ is the total duration of the experiment with $N$ integer. We have verified that the amplitude of the second temporal mode $\eta(x,y,2 \omega)$, for the lowest frequency ($\omega = 2 \pi$ $\mathrm{s}^{-1}$), is around 20\% of the linear mode. This amplitude decreases rapidly as a function of frequency up to $\omega \approx 20$ $\mathrm{s}^{-1}$ where it is smaller than 5\% of the linear mode. As we shall see further in the experimental results, the effect of pinned contact line starts to be important at $\omega>20$ $\mathrm{s}^{-1}$, thus we can consider that our system is dominated by the linear mode. Since in this case the FTP measures the perturbation from a reference already deformed by a meniscus in still water, the transverse profile of the perturbation at the boundary, as we can see in Fig. \ref{fig5pcolor}$a$, becomes curved ($|\partial_y \eta \left( y=\pm W/2 \right) | > 0$) for the pinned edge condition and flat ($\partial_y \eta\left( y=\pm W/2 \right) \approx 0$) for the slipping edge condition. The deformed surface for the case with pinned contact line can be decomposed in the transverse modal basis of eq. \eqref{eq:eta_projection} in order to get the $x$-dependent functions $A_n(x)$. The reconstruction of the transverse profile of an experiment with pinned contact line is shown in Fig. \ref{fig5pcolor}$a$ in dashed line, where the transverse modes decomposition was computed with $n=20$ modes showing good agreement in the whole profile. The surface curvature generated by the wave field in the pinned case can also be observed in Fig. \ref{fig5pcolor}$c$, where the isolines form closed ellipses due to the higher surface deformation in the center of the waveguide. In contrast, in the slipping case the wave field in Fig. \ref{fig5pcolor}$b$ is invariant in the $y$-direction. The difference between both experiments can also be observed in terms of wavelength. Considering that both experiments were measured at the same wave frequency ($\omega=25$ $\mathrm{s}^{-1}$) and with the same channel width ($W=32$ mm), we would not expect any difference in the $x$-direction. However, at a glance they show a different wavelength, a fact that make them with opposite phases after 3 wavelengths.

\begin{figure}[h]
\centering 
	\includegraphics[width=14cm]{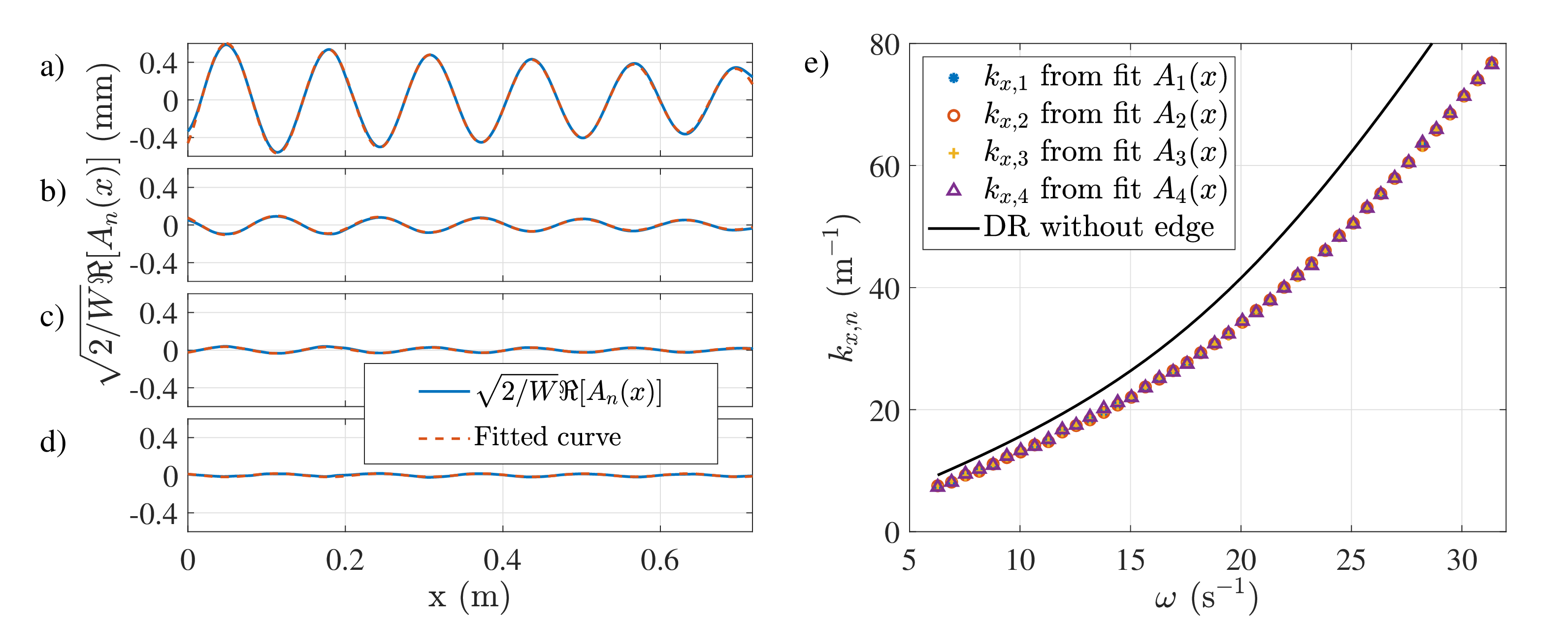}%
 \caption{Longitudinal functions $A_n(x)$ (real part)  and fitted linear wave for the case $W=22$ mm and $\omega=24.5$ $\mathrm{s}^{-1}$. Panels $a)$, $b)$, $c)$, and $d)$ correspond to the functions $A_1(x)$, $A_2(x)$, $A_3(x)$ and $A_4(x)$ respectively. $e)$ Fitted wavenumber $k_{x,n}$ from the longitudinal functions $A_n(x)$ with $n=1,2,3,4$ as a function of frequency.}
	\label{fig6k_fit_graph}
\end{figure}

The functions $A_n(x)$ are fitted with a linear propagating wave using eq. \eqref{eq:tilde_a_n} in order to get the coefficients $a_n$, $r_n$ and more importantly wave vectors $k_{x,n}$. In Figs. \ref{fig6k_fit_graph}$a-d$, for one fixed frequency, we present the real part of the functions $A_n(x)$, with $n=[1,2,3,4]$ and the fitted curves. We observe a very good agreement of the linear approximation and a decreasing amplitude as a function of $n$. In Fig. \ref{fig6k_fit_graph}$e$, varying $\omega$ between experiments, we present the fitted wavenumber $k_{x,n}$ (real part) for $n=1,2,3,4$, showing that, at each frequency, all the functions $A_n(x)$ have the same longitudinal wavenumber ($k_{x,1}=k_{x,2}=k_{x,3}=... =k_{x,n}$). Thus, without loss of generality we can use $k_x=k_{x,1}$ as an input in eq. \eqref{eq:lambda_n} to calculate $\lambda_n$  which is used in the transverse decomposition in eq. \eqref{eq:phi_projection} and further calculations. On the other hand, the imaginary part of $k_{x,n}$ gives the wave spatial damping. As expected, the spatial decay is inversely proportional to the channel with. As a function of $\omega$, the spatial decay is in the range: $\Im(k_{x,1}) \in [0.60,0.90]$ $\mathrm{m}^{-1}$ in the case $W=22$ mm; $\Im(k_{x,1}) \in [0.40,0.68]$ $\mathrm{m}^{-1}$ in the case $W=32$ mm; and $\Im(k_{x,1}) \in [0.35,0.55]$ $\mathrm{m}^{-1}$ in the case $W=42$ mm. We have verified that these values of spatial damping agree with the boundary layer approximation by \cite{hunt1952viscous}. Regarding the amplitude fitted coefficients in the whole frequency spectrum, the incident waves coefficients are in the range $a_1=0.4 \pm 0.1$ mm, and the reflected wave coefficients $r_1$ are smaller than 0.1. This verifies a linear regime with weak reflection where the maximum wave steepness is $k_x a_1<0.04$.

\begin{figure}[h]
\centering 
	\includegraphics[width=14cm]{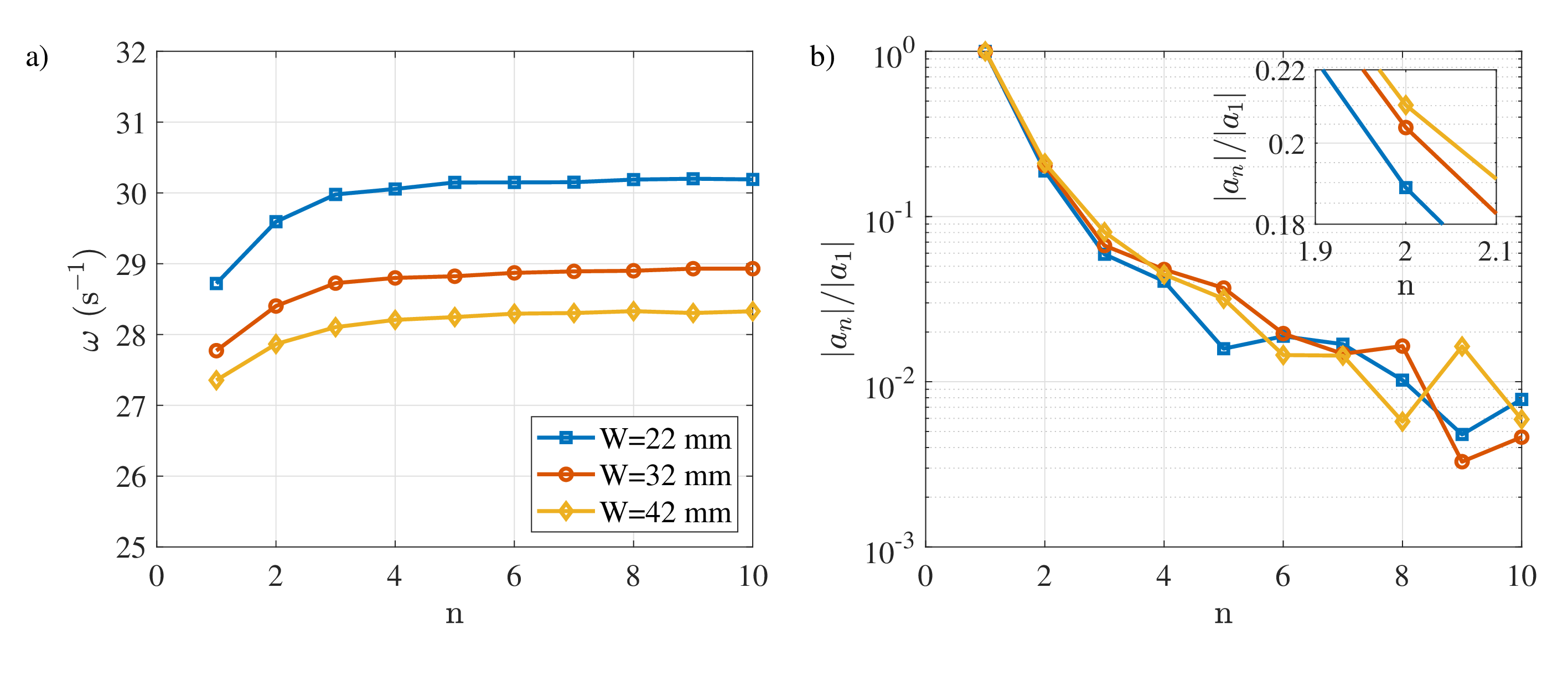}
 \caption{Three experiments at the same wave number $k_x= 71$ $\mathrm{m}^{-1}$ and three different channel widths. $a)$ Frequency $\omega$ as a function of $n$ calculated from eq. \eqref{eq:20}. $b)$ Coefficients $a_n$ obtained from the expansion of eq. \eqref{eq:eta_projection} and the fit of eq \eqref{eq:tilde_a_n}, normalized by the first mode $a_1$. The inset shows a zoom over the mode $n=2$.}
	\label{fig7omegax}
\end{figure}

Having measured the static profile $\eta_s$, fitted the wave number $k_x$ and calculated the transverse decomposition, we can obtain $\omega$ from eq. \ref{eq:20}, which is plotted in Fig. \ref{fig7omegax}$a$ as a function of the number of transverse modes $n$. Here we want to compare three channel widths for a fixed wave number $k_x= 71$ $\mathrm{m}^{-1}$. We can observe small variations and a plateau after 5 modes confirming the accuracy, robustness and convergence of the frequency. Considering that at the same $k_x=71$ $\mathrm{m}^{-1}$ the curvature due to wave steepness is also the same, the difference in frequency observed in Fig. \ref{fig7omegax}$a$ can only be explained due to transverse constraints. In Fig. \ref{fig7omegax}$b$, we show the relative contribution of the fitted coefficients $a_n$ which is rapidly decreasing as a function of $n$. The contribution of the modes $n>8$ is in the order of 1\% of the first mode which is consistent with the convergence of the frequency. Analyzing more in detail the first 3 modes, we observe that at a fixed $k_x$, the narrower is the channel the smaller is the contribution of the modes $n=2$ (as we observe in the inset of Fig. \ref{fig7omegax}$b$) and $n=3$. The small contribution of higher modes ($n=[2,3]$) indicates a more stretched surface produced by the tension between the pinned contact line against the wave induced perturbation. In this case, the transverse profile of the surface perturbation is higher in the center of the channel, due to the proximity of the lateral walls. In contrast, when lateral walls are far from each other, the transverse profile of the surface perturbation tends to be flat (rectangular) with larger contribution of higher modes. 

\begin{figure}[h]
\centering 
	\includegraphics[width=8cm]{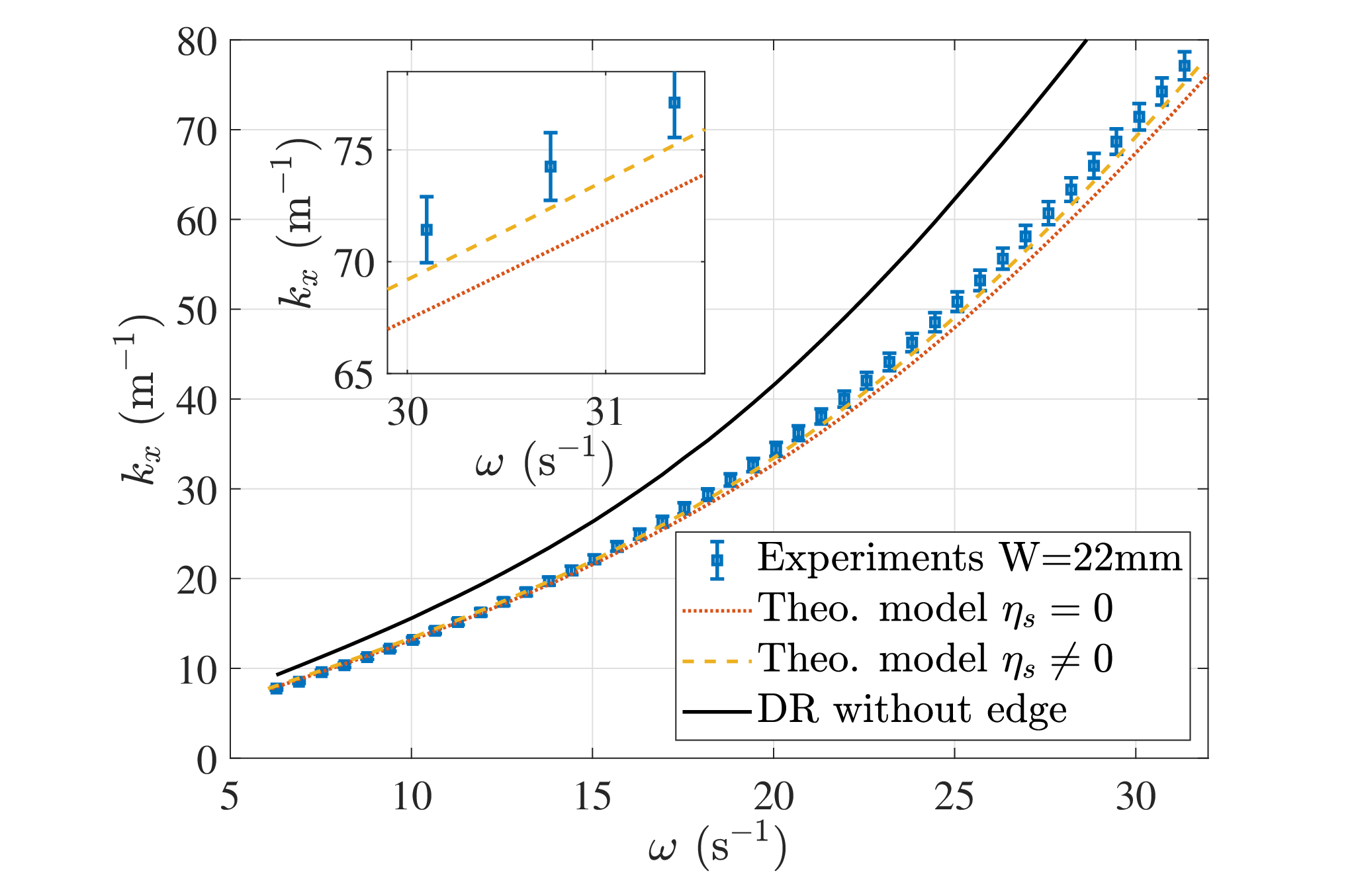}%
 \caption{Theoretical dispersion relation for the case $W=22$ mm considering the relative contribution of measured meniscus $\eta_s$. Dotted line indicates the case $\eta_s=0$, dashed line shows the case $\eta_s \neq 0$ (experimentally measured) and solid line shows the dispersion relation of water waves without edge constraints.}
	\label{fig8_DR_exp_eigen_22mm}
\end{figure}

Next, we have varied $\omega$ between experiments to explore the whole dispersion relation. In order to quantify the contribution of the space-time resolved measurement of the static meniscus and the surface perturbation, we present in Fig. \ref{fig8_DR_exp_eigen_22mm} three curves obtained theoretically, with $n=20$ modes, and compared with the experimentally measured values for the case $W=22$ mm. First, we computed the frequency $\omega$ from eq. \eqref{eq:20} considering a flat static meniscus with $90^{\circ}$ contact angle, that is $\eta_s=0$, and calculating the eigenvectors $b_n$ from the minimization of the system of eqs. \eqref{eq:20}. This curve is plotted in dotted line and corresponds to the lowest estimation of the dispersion relation. The computation is improved in the dashed line when the static meniscus is considered, that is $\eta_s \neq 0$. This computed dispersion relation has the best agreement to the experimental data with a difference smaller than $0.5\%$. In Fig. \ref{fig8_DR_exp_eigen_22mm} errorbars represent the estimation of the error of the experimental measurements. The errors due to accuracy of the instruments are: the water depth with an error of $\Delta h =\pm 1$ mm, the channel width with an error of $\Delta W = \pm 0.1$ mm and the pixel size of the FTP technique with an error of $\Delta dx = \Delta dy= \pm 0.002$ mm.

\begin{figure}[h]
\centering 
	\includegraphics[width=12cm]{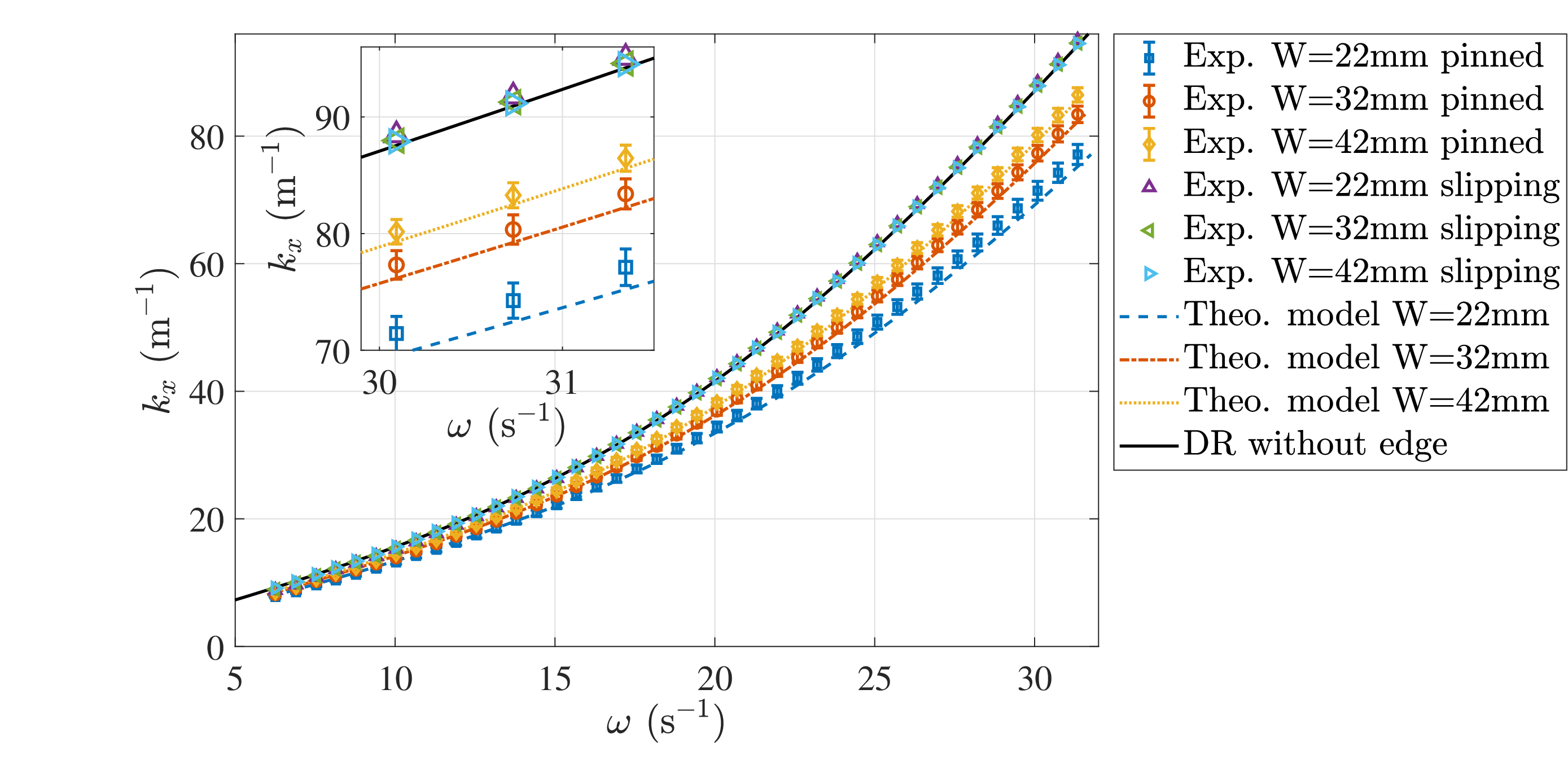}%
 \caption{Dispersion relation from the experimental data and the theoretical model. Symbols: experiments with pinned and slipping contact line for the three channel widths: $W=[22,32,42]$ mm; Solid line shows the dispersion relation from eq. \eqref{eq:DR}; Dashed, dash-dot ant dot lines show the theoretical model from eq. \eqref{eq:20} for the three channel widths.}
	\label{fig9_graph_DR_omega_diff}
\end{figure}

Eventually, we present in Fig. \ref{fig9_graph_DR_omega_diff} the experimentally measured dispersion relation for the three channel widths with two different edge conditions: pinned contact line (bare wall) and slipping contact line (wire mesh). As we observe, with pinned contact line, the dispersion relation is shifted down when smaller is the channel width. In contrast, the condition of slipping contact line does not change the dispersion relation, independently of the channel width. Thus, for a fixed frequency, the slipping contact line wavenumber is always higher than the pinned contact line wavenumber. For example, at the maximum explored frequency, $\omega = 31.4$ $\mathrm{s}^{-1}$, the difference in wavenumber between the slipping and the pinned contact line conditions is $17.9$, $11.2$ and $8.1$ $\mathrm{k}^{-1}$ for the channel widths $W=22$, $32$ and $42$ mm respectively. In the other direction, for a fixed wavenumber, the pinned frequency is always larger than the slipping frequency. The theoretical dispersion relation for each channel width was calculated with $n=20$ transverse modes and takes into account the experimentally measured static meniscus $\eta_s \neq 0$.

As we observe in Fig. \ref{fig9_graph_DR_omega_diff}, the agreement between the experiments and the theoretical model at high frequency ($\omega>25$ $\mathrm{s}^{-1}$) is worse for small channel width. The limited spatial resolution of the optical technique (FTP) makes that the number of points in the $y$-direction decreases with the channel width. Therefore, the resolution and accuracy in the measurement of the static meniscus $\eta_s$ and the surface perturbation $\eta(x,y)$ is necessarily lower in a very narrow channel. This problem rises from the experimental compromise between having a high spatial resolution or having a large number of spatial wave periods. In the first case, we can have a better resolution for the measurement of the static profile and the surface perturbation. However, errorbars in the measurements of $k_x$ would be larger due to limited number of spatial wave periods in the $x$-direction. In the second case, which is the case that we have chosen in this work, we have an accurate measurement of the dispersion relation but with a limited accuracy in the estimation of the influence of transverse constraints.
In this article we have preferred to show an accurate dispersion relation, which we consider an asset of our space-time resolved experiments. Despite this fact, the theoretical model is still in the errorbar of the dispersion relation, confirming the pertinence of the linear approximation.

\section{Conclusion}
In this article we report on the FTP measurements of the propagation of surface gravity-capillary waves in a rectangular channel. Two different boundary condition were tested: bare walls cleaned previously with ethanol setting hydrophobic conditions which pinned the contact line, and walls covered with a wire mesh setting wet conditions which allows an easy slipping of the contact line. 

We have varied the channel width and measured the dispersion relation for both boundary conditions. We verified experimentally that a slipping contact line makes the dispersion relation independent of the channel width. In contrast, a pinned contact line modifies the dispersion relation up to 20\% lower in wavenumber when the channel width is divided by two. In Fig. \ref{fig9_graph_DR_omega_diff} we have shown an experimental confirmation of the model developed by \cite{shankar2007frequencies} where we have in addition inserted experimental measurements of the static meniscus and surface perturbation. Regarding the static meniscus, when we compare its contribution with respect to a flat still profile, the precision of the dispersion relation is improved around 4\%. These experimental results reveal that the pinned contact line exerts a force in the opposite direction of the wave displacement. This force, acts as an additional restoring force increasing, together with the gravity and the surface tension, the phase velocity of the wave. On the other hand, we have observed from the experiments, that the slipping contact line follows accurately the dispersion relation of gravity capillary waves, being the influence of the pinned contact line the only source of disagreement between both experimental series. Besides, we observed as expected, that the influence of the pinned contact line and its restoring force is local, that is, the larger is the fluid domain, the lower is the influence in the dispersion relationship. 

The advantage of the space-time resolved measurement is a robust statistics, where we were able to compute the natural frequency by using all the points in the longitudinal direction (around 1000 points) to fit the functions $A_n(x)$, get the transverse coefficients $a_n$ and the wavenumbers $k_x$. That allows us to shift the natural frequencies ($\omega=\omega_1$ from eq. \eqref{eq:20}) closer to the experimental data. This verifies our hypothesis of improving the estimation of the dispersion relation by measuring precisely the actual surface deformation.

The present experiments can be easily applied to different geometries like cylindrical containers. The measure of the surface deformation in each point of the domain gives a statistically strong way to apply \textit{weak} solutions to this type of  problems where analytic solutions are difficult. An interesting continuation of this work may include the measure of the nonlinear effects of a sliding contact line in non-wetting (hydrophobic) or partial wetting conditions, by increasing the wave amplitude beyond the meniscus height to force the sliding of the contact line. Besides, different types of surfaces can also be tested, like porous, granular or inclined walls. The measurement of the contact angle and the theoretical relation of the small scale with the surface wave scale is also a relevant work for a comprehensive analysis of the problem.

\vspace{0.5cm}
E.M. acknowledges the support of CONICYT (ANID) Becas Chile Doctorado.

\bibliographystyle{unsrt}


\appendix
\section{Surface curvature}
\label{appendixA}
The general curvature of the water surface is
\begin{equation}
\hat{c}=\frac{\left(1+(\partial_y \eta)^2 \right) \partial_{xx}\eta -2 \partial_x \eta \partial_y \eta \partial_{xy} \eta + \left(1+(\partial_x \eta)^2 \right) \partial_{yy}\eta}{\left(1+(\partial_x \eta)^2+ (\partial_y \eta)^2\right)^{3/2}}
\label{eq:kappa}
\end{equation}
where we replace the expansion of eq. \eqref{eq:eta expansion} and the projection of eq. \eqref{eq:eta_projection} to obtain the linearized perturbation of the curvature
\begin{equation}
\begin{split}
c(x,y)=\frac{-3\partial_y\eta_s c_s \partial_y\eta}{\left(1+(\partial_y \eta_s)^2\right)}
+ \frac{\partial_{yy}\eta}{\left(1+(\partial_y \eta_s)^2\right)^{3/2}} \\ + \frac{\partial_{xx}\eta}{\left(1+(\partial_y \eta_s)^2\right)^{1/2}} 
\label{eq:kappa_hat}
\end{split}
\end{equation}
where the derivatives of the surface perturbation are
\begin{align}
\partial_y \eta=&-\nu_n a_n \sin \nu_n y \\
\partial_{yy}\eta=&-\nu_n^2 a_n \cos \nu_n y \\
\partial_{xx}\eta=& - k_x^2 a_n \cos \nu_n y 
\end{align}

\section{Integrals}
\label{appendixB}
Integrals over the interval $y=[0,1/2]$ of the projection of eqs. \eqref{eq:17a} and \eqref{eq:17b} onto the transverse basis $\cos \nu_i y$ 
\begin{equation}
\begin{split}
\gamma_{i,n}=\int_0^{1/2} \left[ \lambda_n \cos \left( k_ny \right) \frac{\sinh \left[ \lambda_n(\eta_s(y)+h) \right]}{\sinh \left( \lambda_n h \right)} \right. \\ \left. +  k_n\partial_y \eta_s \sin \left( k_ny \right) \frac{\cosh \left[ \lambda_n(\eta_s(x)+h) \right] }{\sinh \left( \lambda_nh \right) } \right] \cos \left( \nu_i y \right) dy
\end{split}
\end{equation}
\begin{equation}
\delta_{i,n}=\int_0^{1/2}  \cos \left( k_ny \right) \cos \left( \nu_i y \right) \frac{\cosh \left[ \lambda_n(\eta_s(y)+h) \right] }{\sinh \left( \lambda_nh \right)} dy
\end{equation}
\begin{equation}
\xi_{i,n}=\frac{1}{a_n}\int_0^{1/2} \left[ a_n \cos \left( \nu_n y \right) -\beta \hat{c}(x,y) \right] \cos \left( \nu_i y \right) dy
\end{equation}

%
%
%

\end{document}